# LinBFT: Linear-Communication Byzantine Fault Tolerance for Public Blockchains

Yin Yang

yin@yang.net

*Abstract*— **This paper presents LinBFT, a novel Byzantine fault tolerance (BFT) protocol for blockchain systems that achieves amortized $O(n)$ communication volume per block under reasonable conditions (where *n* is the number of participants), while satisfying determinist guarantees on safety and liveness. This significantly improves previous results, which either incurs quadratic communication complexity, or only satisfies safety in a probabilistic sense. LinBFT is based on the popular PBFT protocol, and cuts down its $O(n^4)$ complexity with three tricks, each by $O(n)$: linear view change, threshold signatures, and verifiable random functions. All three are known, i.e., the solutions are right in front of our eyes, and yet LinBFT is the first $O(n)$ solution with deterministic security guarantees.**

**Further, LinBFT also addresses issues that are specific to permission-less, public blockchain systems, such as anonymous participants without a public-key infrastructure, proof-of-stake with slashing, rotating leader, and a dynamic participant set. In addition, LinBFT contains no proof-of-work module, reaches consensus for every block, and tolerates changing honesty of the participants for different blocks.**

## I. INTRODUCTION

The blockchain technology, pioneered by Bitcoin [23], promises to revolutionize finance with a secure, decentralized and trustless protocol for processing transactions, which include money transfers and smart contracts [12]. Many blockchain systems nowadays, however, suffer from poor scalability and slow confirmations, and consume vast amounts of energy [10]. At the heart of the problem is the widely-used *proof-of-work* (*PoW*) census mechanism, in which special power nodes, called *miners*, compete to solve cryptographic puzzles in order to gain the privilege of confirming transactions. Aside from scalability, latency and sustainability issues, PoW also inevitably *forks* [16], i.e., different nodes may have conflicting views of the blockchain. Consequently, PoW usually requires a separate mechanism, e.g., waiting for a number of confirmations, to ensure transaction finality in a probabilistic sense, which introduces uncertainty and exacerbates the latency problem. In addition, PoW chains are vulnerable to the so-called "51% attack", in which an adversary with overwhelming computation power (e.g., an ASIC maker, or a botnet) deliberately forks the blockchain to invalidate previously confirmed transactions. 51% attacks are becoming increasingly common: notably, a recent one compromised Bitcoin Gold[1], which involves a relatively large number of nodes, and was designed to be ASIC-resistant.

An attractive alternative to PoW is *Byzantine Fault Tolerance* (*BFT*), which involves negligible computations, guarantees no fork, and promises near-instant block finality. BFT has been well-studied in the context of distributed systems. However, there is much confusion on how to apply BFT in a blockchain setting, which is quite different from traditional distributed systems. Most importantly, BFT has been perceived to be a communication-heavy protocol. In particular, there is a long-standing myth that BFT is not scalable to the number of participants *n*, since most existing solutions incur the transmission of $\Omega(n^2)$ messages, even under favorable network conditions. As a result, existing BFT chains involve very few nodes (e.g., 21 in [14]), which can be elected delegates [14], PoW winners [21], or a random sample set obtained through cryptographic sortition [18][16]. As discussed in [18][16], having fixed delegates defeats decentralization, and PoW introduces uncertainty in the participant set due to forks. Sortition-based BFT, on the other hand, only provides probabilistic guarantees on safety (i.e., no fork); further, its probability of failure is only small enough with a large sample (minimum a few hundred nodes [18]), which might be already beyond the capacity of currently deployed BFT chains (e.g., 21 nodes in EOS [14], and up to 16 in Hyperledger Fabric [11]).

This paper dispels the myth with a novel BFT protocol, called LinBFT, that achieves amortized $O(n)$ transmission cost per block with deterministic guarantees on safety and liveness, under reasonable assumptions, elaborated in Section III. This is clearly optimal, in the sense that disseminating a block already takes $\Omega(n)$ transmissions[2]. LinBFT is based on PBFT [5], an $O(n^4)$ solution, and reduces its communication costs with three tricks, each by $O(n)$: linear view change [1], threshold signatures [18], and leader selection via verifiable random functions (VRFs) [16]. All three exist in previous methods, and their combination is simple; yet, to the author's knowledge, LinBFT is the first amortized-$O(n)$ BFT protocol with deterministic security guarantees.

The proposed protocol is designed for permission-less, public blockchains, for which there are additional challenges not commonly encountered in other BFT settings. For instance, participants can be anonymous, without a centralized public key infrastructure (PKI). Consequently, nodes must exchange public keys among themselves, and participate in a distributed key generation (DKG) protocol required to create threshold signatures, both of which are communication-heavy processes. Meanwhile, LinBFT is compatible with proof-of-stake (PoS), which counters Sybil attacks and deters dishonest behavior through slashing, as described in Section II-B. Additionally, LinBFT has the following desirable features:

- No PoW. This avoids uncertainty due to inherent forks in PoW, be it in the participant set, leader selection, or block confirmation. More importantly, this eliminates 51% attacks, which are a serious risk for new, smaller PoW chains.
- Per-block consensus. There is consensus for each individual block, rather than for a group of blocks, e.g., in Casper [3]. This limits the power of the block proposer, and, thus, mitigates selfish mining.
- Rotating leader. The proposed protocol changes the leader (i.e., block proposer) for every block, which reduces the risk of denial-of-service attacks on the leader.
- Changing honesty. In LinBFT, a participant can be honest for one block, and malicious for another (e.g., one containing a transaction of interest to the participant), as long as over 2/3 of all participants are honest for each block. In other words, it is possible that every participant is malicious at some point, and yet the blockchain remains secure at all times.
- Dynamic participant set. LinBFT allows nodes to join and leave the protocol at the beginning of epochs (explained in Section IV-C. As a result, it is possible that different blocks are verified by completely different sets of nodes.

---

[1] *https://www.coindesk.com/blockchains-feared-51-attack-now-becoming-regular/*

[2] Although it is impossible to obtain sublinear transmission volume, a method can still have sublinear communication *time* in an ideal environment. LinBFT achieves best-case $O(\log n)$ time through speculative execution, described in Section IV-A.

## II. PRELIMINARIES

### A. PBFT with Linear View Change

The basis of LinBFT is PBFT [5], a classic BFT protocol that requires $O(n^2)$ transmissions in its best case, and $O(n^4)$ in the worst case. PBFT can be used to obtain consensus on a block in a blockchain setting (e.g., implemented in [2]), as follows. Assume that there are $n = 3f+1$ nodes in total, among which $f$ are malicious. The protocol involves one or more *rounds*, each of which has a *leader* node, which can be chosen, e.g., in a round-robin manner. The leader in the $i$-th round (denoted by $L_i$) proposes a block $b_i$, and broadcasts its hash value $h_i$ in a *Prepare* message. Upon receiving such a *Prepare*, a node that is either not locked or already locked on the same digest $h_i$ (locking is elaborated shortly) responds by broadcasting *Prepare* messages on $h_i$. Once a node receives $2f+1$ *Prepare*'s on $h_i$, it assembles them into a *commit certificate* (*CC*), locks on $h_i$, and broadcasts *Commit* messages about the CC. A node who receives $2f+1$ *Commit*'s on the CC of $h_i$ is ready to finalize block $b_i$, after verifying all transactions in the block.

**Linear View Change.** When consensus cannot be reached in a round $i$ within a given timespan, the protocol enters a new round $i+1$ with a different leader $L_{i+1}$, which is called a *view change*. Possible causes for a view change include a malicious leader (who can simply remain silent throughout the round), or message losses due to network failures. The original PBFT protocol involves $O(n^3)$ transmissions per view change. A recent algorithm called *linear view change* (*LVC*) reduces the cost to $O(n^2)$ [1]. Without network failures, there can be at most $f$ view changes due to malicious leaders, leading to $O(n^3)$ overall communication complexity with LVC, or $O(n^4)$ in the original PBFT.

LVC works as follows. At the beginning of a new round $L_{i+1}$, each node sends a *NewView* message to $L_{i+1}$, along with the commit certificate it is locked to, if any. $L_{i+1}$ then broadcasts the CC with the highest round number among all collected CCs, along with the hash $h_{i+1}$ of its proposed block $b_{i+1}$. A locked node unlocks, whenever it receives a CC with a higher round number compared to the one it is currently locked to [1]. Since a CC contains $2f+1$ *Prepare* messages, it has size $O(n)$; thus, collecting and broadcasting a CC cost $O(n^2)$ transmissions.

**Safety, liveness, and partial synchrony.** Two fundamental requirements for a BFT protocol are *safety* (i.e., once the protocol finishes, all honest nodes agree on the same block) and *liveness* (the protocol eventually terminates). PBFT satisfies safety as long as $n \geq 3f+1$, even in the presence of network failures. It satisfies liveness in a *partially synchronous* network [9], in which there can be arbitrary network failures until a particular time point (not known beforehand), after which the network becomes *synchronous*. Here, network synchrony is defined as the condition that any message sent from one honest node to another is guaranteed to be delivered correctly within $\Delta$ time, where $\Delta$ only depends on the length of the message.

### B. Threshold Signatures

The communication cost of PBFT can be reduced by another factor of $O(n)$ with *threshold signatures*, each of which combines multiple signatures into one, and, thus, reduces the size of a commit certificate from $O(n)$ to $O(1)$ [1]. Formally, an $(n, t)$-threshold signature on a message $m$ is an aggregate signature that passes verification if and only if no less than $t+1$ participants sign $m$. Accordingly, a CC in PBFT can be implemented using a threshold signature with $t = 2f$. Note that the verifier of a threshold signature does not need to know the identifies of the signers, which eliminates the need to include a size-$O(n)$ identity vector in the aggregate signature, e.g., as is done in [25]. LinBFT employs a popular implementation of threshold signature based on the BLS signature scheme [4].

**Setup costs of threshold signatures.** Note that *threshold signature is not free lunch*, a fact that is sometimes ignored in the literature. In particular, a threshold signature scheme requires special, correlated public / private key pairs. Generating such key pairs in a decentralized setting requires a distributed key generation (DKG) protocol, which is communication-heavy. For example, the Joint-Feldman algorithm [17], e.g., used in Dfinity [18], requires each node to broadcast the coefficients of an order-$t$ polynomial. This incurs $O(n^2t)$ transmissions, which is $O(n^3)$ when $t = 2f = O(n)$. Further, the original proposal in [17] requires network synchrony, which is relaxed in later work, e.g., [20], at the expense of even higher worst-case communication overhead. LinBFT employs the DKG solution in [7], which requires $O(n\ polylog\ n)$ communications with a given participant set, and provides a probabilistic guarantee on the correctness of the generated key pairs. This method requires a common source of randomness among the participants, which is conveniently provided by LinBFT through verifiable hash functions, detailed next.

### C. Verifiable Hash Function with Consensus

A *verifiable hash function* (*VRF*) is a pseudo-random generator whose output is verifiable (i.e., on whether a given number is indeed the output of the VRF), random, uniformly distributed, and unpredictable beforehand. A simple VRF (e.g., used in Algorand [16]) under the *random oracle* model (i.e., there exists an ideal hash function $H$ whose outputs are random and uniform), is $H(s)$, where $H$ is the ideal hash function and $s$ is a signature that satisfies *uniqueness*, i.e., there is a unique signature for a given message and a private key. The BLS signature scheme [4], for instance, satisfies this property. The output of $H(s)$ is clearly random and uniform, due to the random oracle assumption. Meanwhile, given the source message $m$ (from which $s$ is obtained) and $s$'s corresponding public key, one can verify that a given value is indeed the result of $H(s)$. Further, the function's output is unpredictable, since it is infeasible to obtain $s$ without knowledge of its secret key.

LinBFT additionally requires the VRF to reach consensus among a supermajority (i.e., $2f+1$) of nodes. This is done by hashing a BLS threshold signature $ts$, i.e., $H(ts)$, which is used, e.g., in Dfinity [18]. In the BFT setting, this ensures that the VRF remains unpredictable in the presence of $f$ malicious nodes.

### III. PROBLEM SETTING

### A. Cryptographic Foundation

LinBFT operates in the random oracle model of cryptography, with an ideal hash function $H$ explained in Section II-C. In practice, $H$ can be approximated with a cryptographical one, such as SHA-3 [8]. Additionally, since LinBFT employs BLS signatures [4], it inherits the latter's underlying assumptions, e.g., the hardness of discrete logarithm on an elliptic curve with pairing.

### B. Proof-of-Stake

Proof-of-stake (PoS) is a common technique to counter *Sybil attacks*, i.e., one single person or entity pretends to be many participants by registering numerous accounts in the system. In the presence of such attacks, it is no longer appropriate to define BFT's honest supermajority requirement based on the number of nodes, since multiple nodes can belong to the same entity.

PoS addresses this problem by mandating that each participant deposit to a special account a certain amount of money (call its *stake*) in the form of cryptocurrency tokens of the blockchain[3]. The stake can only be withdrawn after the participant quits the protocol. Clearly, under PoS, the number of accounts that a single person or entity can register is limited by its financial resources. Thus, the honest supermajority requirement of BFT can be defined as over 2/3 of the staked money is honest. With every participant staking the same amount, this requirement can be stated in the familiar form of $n \geq 3f+1$. As pointed out in [16], the honest-money condition is comparable to PoW's assumption that most of the computation power in the system is honest, since money can buy computational power.

**Slashing.** In some PoS systems such as Casper [3], a participant that is caught cheating gets *slashed*, meaning that the cheater is kicked out of the participant set with its stake confiscated. It is argued in [3] that removing the cheater's stake from the system altogether (i.e., the staked tokens go to a "black hole" that cannot be transferred to anyone) reduces the total token supply of the blockchain, and, thus, pushes up the price for each token. Intuitively, this compensates for the damage done by the dishonest participant.

### C. Adversary Model

Out of the $n$ participants of LinBFT, $f$ (satisfying that $n \geq 3f+1$) can be *malicious* for each execution of the protocol, which can misbehave in arbitrary ways[4]. The remaining participants are *honest*, who strictly follow the protocol. Since the malicious nodes can collude, one can view them as corrupted and controlled by a single mastermind, referred to as "the adversary". LinBFT assumes that the adversary is *static* and *rushing* (i.e., it has to choose the $f$ nodes to corrupt before a protocol run), rather than *adaptive* (which can instantly compromise any node at any time). Note that for a fast protocol, the adaptive-adversary assumption might be overly pessimistic. Further, the adversary is assumed to take time to compromise nodes. In LinBFT, this time is assumed to be no less than the duration from the point when a block at a certain height (say $l$) is first finalized by a node, to the beginning of the protocol for the next block height $l+1$. The reason for this is explained later in Section IV-B.

Since the blockchain setting contains an infinite number of transactions split into blocks, it is important to clarify the timespan over which $f$ is defined. As an extreme case, if the adversary is limited to corrupting $f$ nodes for the entire lifetime of the blockchain, then, once we catch those $f$ bad actors, we no longer need BFT since all remaining nodes are honest. In LinBFT, the honesty of a participant is allowed to change from block to block, e.g., an honest participant may become malicious when it encounters a transaction that strongly motivates it to cheat. Accordingly, over multiple blocks, it is possible that every participant is malicious at some point, and yet the system remains secure as long as $n \geq 3f+1$ holds for each block. In this sense, LinBFT provides stronger security compared to systems (e.g. Casper [3]) that run consensus once for multiple blocks. Meanwhile, the changing honesty assumption makes it necessary to place the condition that the adversary cannot carry over knowledge of private keys of malicious nodes from one execution of the protocol to another.

---

[3] PoS can also be implemented with different participants staking different amount of tokens, and having influence proportional to their respective stakes. For ease of presentation, the rest of the paper assumes a fixed amount of stake for each participant, i.e., all participants are equal, and multiple participants can be controlled by the same entity.

[4] Some protocols such as SBFT [15] distinguish malicious nodes that actively send falsified messages from fail-stopped ones that crash temporarily, in order to obtain higher fault-tolerance capacity. LinBFT does not make this distinction since it is hard to estimate the proportion of each type in practice.

### D. Communication Model

Following common practice in the literature, LinBFT assumes that the network is partially synchronous [9], i.e., after an unknown future time, any message between two honest nodes is delivered within $\Delta$ time, as explained in Section II-A. This model also captures the more common situation that periods of synchrony and asynchrony interleave, and there are sufficiently long periods of synchrony that allow the protocol to finish [1].

It is worth pointing out that the parameter $\Delta$ above does not take into account the network topology, or the total amount of network traffic. For instance, an all-to-all broadcast, which clearly involves $\Omega(n^2)$ messages, can be said to take $O(n)$ time, since every node sends/receives $n-1$ messages, each of which takes constant time, i.e., up to $\Delta$. This analysis is not valid when the network has a saturated critical link, e.g., when half of the nodes reside in America and the other half in China. In this case, since $\Omega(n^2)$ traffic pass through this critical link with limited bandwidth, the total time is no less than $\Omega(n^2)$. For this reason, LinBFT focuses on minimizing communication volume rather than time, as discussed in the next subsection.

Lastly, LinBFT involves both broadcast and single-cast communications. Thus, it is not compatible with a gossip-only setting, e.g., [6][16]. In reality, single-casts are ubiquitous, e.g., a browser connecting to a web server, and gossip-only settings might be overly restrictive.

### E. Security and Performance Objectives

**Security objectives.** As a BFT protocol, LinBFT must satisfy *safety* (i.e., no fork) and *liveness* (the protocol eventually terminates) when the number of malicious nodes is less than a third of all participants (i.e., $n \geq 3f+1$), explained in Section II-A. Specifically, in a partially synchronous network described in the previous subsection, LinBFT satisfies both safety and liveness deterministically, with zero chance of failure. Further, LinBFT also satisfies deterministic safety even when the network is asynchronous, i.e., it never forks.

Additionally, LinBFT also aims to maintain decentralization by limiting the power of individual nodes, and to mitigate the problem of *selfish mining*, in which a block producer prioritizes certain transactions over others. This is done by changing the leader (who proposes new blocks) for every block height and every round within the same height, and choosing the leader uniformly, elaborated in the next section.

**Performance objectives.** LinBFT aims to achieve amortized $O(n)$ worst-case transmission volume per block, unless with negligible probability, subject to the following conditions. First, the communication complexity is calculated based on messages required to complete the protocol, e.g., invalid messages sent by malicious nodes do not count. Second, since LinBFT does not guarantee progress during periods of asynchrony (which is impossible according to [13]), and such a period can be arbitrarily long, it is necessary to only consider transmissions during network synchrony. These assumptions are implicitly made in most BFT protocol analyses.

Third, since there is a large setup cost for generating and exchanging keys, the participant set cannot change too frequently. For instance, exchanging public keys and IP addresses between each pair of nodes already take $O(n^2)$ transmissions. In the literature, such costs are often hidden by assuming the existence of a centralized public-key infrastructure (PKI), which is not practical in a permission-less, public blockchain setting. To amortize the setup costs, LinBFT divides time into *epochs* of length $O(n)$, and nodes can only join or leave at the beginning of each epoch, which is detailed later in Section IV-C.

Lastly, slashing (Section III-B) can also lead to a high communication cost. In an extreme case that all $f = O(n)$ nodes are slashed in the same protocol run, the remaining $O(n)$ honest nodes need to know the cheaters' identities, leading to $O(n^2)$ transmissions. Hence, linear complexity can only be obtained with at most a constant number of slashes per protocol run. More slashes lead to a higher communication cost (quadratic in the worst case), but do not affect safety or liveness. Note that the time loss due to slashing is compensated financially by confiscating the cheaters' stakes, as explained in Section II-B.

**Communication volume vs. time.** LinBFT focuses on minimizing total transmission volume rather than time, since (i) the latter varies depending on the network topology and traffic conditions and (ii) time complexity is no better than transmission complexity, when there exist saturated critical links, as explained in Section III-D. Nonetheless, in the ideal case with a synchronous network and no malicious node, LinBFT obtains $O(\log n)$ communication time through speculative execution, elaborated in the next section.

## IV. LINBFT

LinBFT is executed once for each block height $l$, and only moves on to the next height $l+1$ when a supermajority, i.e., $2f+1$ nodes, reach consensus on a block to be added to the blockchain at height $l$. Hence, it suffices to focus on the protocol run on one particular height $l$. Meanwhile, inside a protocol run, there may be one or more rounds, each of which has a leader, denoted as $L_i$ for the $i$-th round. In the following, Section IV-A describes the protocol run in the ordinary case, where no node misbehaves, and the network is synchronous. Further, the ordinary case assumes that necessary infrastructure is already in place, such as public / private keys for threshold signatures. Section IV-B deals with malicious nodes, and present the view change and leader rotation schemes. Section IV-C handles changes in the participation set. Section IV-D summarizes the protocol and provides security and performance analyses.

### A. Ordinary Case

Similar to PBFT, LinBFT finishes the ordinary case in a single round (i.e., round 1), in three steps: *Preprepare*, *Prepare* and *Commit*. As in PBFT (described in Section II-A), in the first step, the leader (i.e., $L_1$) broadcasts a *Prepare* command containing the hash digest $h_1$ of its proposed block $b_1$. Upon receiving such a message, a node (say, node $u$) needs to notify all others on its acceptance of $h_1$ in the form of a signed *Prepare* message. However, unlike in PBFT, in LinBFT $u$ cannot simply broadcast its *Prepare* on $h_1$, since an all-to-all broadcast would cost $O(n^2)$ transmissions.

As in PBFT, the purpose of the *Prepare* messages is to ensure that in each round, at most one block hash receives *Prepare* votes from a quorum, i.e., $2f+1$ nodes; only this block hash proceeds to the *Commit* step. LinBFT applies a standard trick (e.g., used in [15][19]) with a *collector* that collects and aggregates *Prepare* votes from all nodes. A natural choice for the collector is the leader $L_1$. In the ordinary case, $L_1$ collects *Prepare*'s from all $n$ nodes on $h_1$. Then, $L_1$ derives an ($n$, $t = 2f+1$)-threshold signature $ts(h_1)$ on $h_1$ from these *Prepare* messages, and creates a commit certificate $CC_1$, as follows.

$$CC_i = <i, h_i, ts(h_i)>. \quad (1)$$

The size of the above commit certificate is constant, since a threshold signature is of constant size (refer to Section II-B). After that, $L_1$ broadcasts $CC_1$ to all nodes, each of which responds with a *Commit* message with its signature on $CC_1$. Again, LinBFT uses a collector to avoid all-to-all broadcasts of *Commit*'s, and the collector's role is naturally fulfilled by the leader, $L_1$. In particular, $L_1$ collects *Commit*'s from all $n$ nodes, derives an ($n$, $2f+1$)-threshold signature $ts(CC_1)$ on $CC_1$, and broadcasts $ts(CC_1)$. Once a node receives $ts(CC_1)$, it finalizes block $b_1$ (i.e., by adding it to the local copy of the blockchain) after verifying its transactions[5]. This completes the protocol.

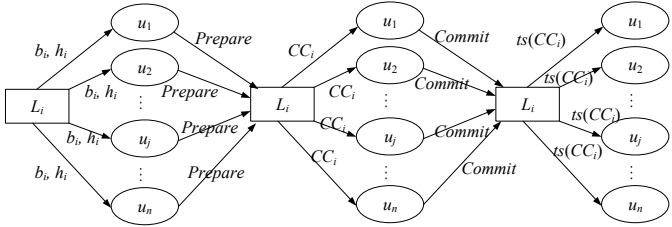

Fig. 1 Example that a leader $L_i$ serves as a collector for *Prepare* and *Commit* messages, computes threshold signatures, and re-broadcasts certificates.

**Speculative execution in $O(\log n)$ time.** In the ordinary case, all $n$ nodes sign $h_1$ and $CC_1$, meaning that there is no ambiguity on the signers. Thus, it suffices to use multi-signatures of all nodes instead of threshold signatures. This makes it possible to apply the CoSi protocol [25], which organizes all nodes into a balanced tree, each of which collects and aggregates (multi-)signatures from its child nodes. According to [25], CoSi achieves $O(\log n)$ time under the synchronous network assumption, i.e., each constant-sized message is delivered in constant time. The total transmission volume remains $O(n)$. Note that this (optional) CoSi run is speculative: if any node fails to collect valid signatures from all its children, it notifies the leader $L_1$, and the protocol then falls back to threshold signatures.

## B. View Change and Leader Rotation

With the presence of malicious nodes, a round can fail to reach consensus for a variety of reasons, e.g., a malicious leader can simply remain silent indefinitely[6]. To ensure liveness, each node sets a timer for every round. When the timer expires, a *view change* is triggered, and the protocol enters a new round, say, round $i+1$. LinBFT handles a view change with the LVC algorithm [1], explained in Section II-A. Specifically, the leader of the new round (i.e., $L_{i+1}$) now serves as the collector for the *NewView*, *Prepare* and *Commit* messages. Since the size of a commit certificate is constant according to Equation (1), the total transmission cost of a view change is $O(n)$.

**Random leader selection.** In PBFT, there can be a cascading sequence of $f$ faulty leaders, leading to $f+1 = O(n)$ rounds. LinBFT avoids this situation by selecting leaders randomly, using a VRF (refer to Section II-C). With random leaders, the probability of having a sequence of malicious leaders diminishes exponentially with the length of the sequence. Specifically, with $f < n/3$ malicious nodes, having a sequence of $x$ malicious leaders has probability smaller than $1/3^x$. Therefore, the probability that the next leader is malicious becomes negligible (i.e., smaller than a given $\rho$) after a constant number ($x > -\log_3 \rho$) of leader changes. In practice, a common choice of $\rho$ is $10^{-18}$, whose inverse is larger than the total number of seconds since the beginning of the universe [16].

Random leader selection requires a common source of randomness among the nodes. In LinBFT, this is provided by a VRF on the threshold signature that finalizes the previous block at height $l–1$. In particular, we have:

$$L_i = H(ts(CC^{l-1}) \mid i) \bmod n, \quad (2)$$

where $CC^{l-1}$ is the last commit certificate at the previous block height $l-1$, and $ts(CC^{l-1})$ is the threshold signature indicating that $2f+1$ participants have signed $CC^{l-1}$. In LinBFT, $ts(CC^{l-1})$ authorizes a node to finalize its corresponding block (after verifying its transactions), which is first generated by the leader of the last round at height $l-1$. Based on the assumption that the adversary takes time to corrupt nodes, which equals the duration from the creation of $ts(CC^{l-1})$ at the last leader and the beginning of the protocol at height $l$, the output of the VRF above is unpredictable to the adversary beforehand. Meanwhile, the VRF is clearly known to all nodes that have finalized at height $l-1$, which is a necessary condition for entering the protocol for height $l$. An example is shown in Fig. 2.

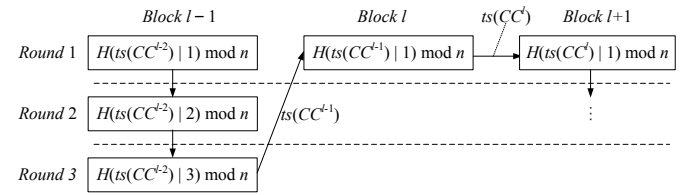

Fig. 2 Example random leader selection for three consecutive blocks.

---

[5] Transactions in a block $b_i$ can be verified either before sending *Prepare* on its hash $h_i$, or before finalizing the block. The latter enables block transmission to be done in parallel to the protocol messages.

[6] Other causes of a view change include a malicious leader proposes multiple blocks or one containing invalid transactions, or a non-leader node proposes a block. Such behaviors are punished by slashing.

With random leaders, the number of view changes in a synchronized network becomes $O(1)$, unless with negligible probability. Since each view change takes $O(n)$ transmissions, the total communication cost for all view changes is still $O(n)$, unless with negligible probability.

Finally, note that Equation (2) chooses a leader by random sampling with replacement. Thus, theoretically, it is possible that it always selects a malicious leader, though with exponentially diminishing probability as the process goes on. As a result, the liveness guarantee of the protocol becomes probabilistic rather than deterministic. To address this issue, it suffices to modify the above leader selection algorithm by generating a random permutation of the $n$ nodes using the common random source $H(ts(CC_{l-1}))$, and choosing leaders round-robin according to this permutation.

### C. Handling Changes in the Participant Set

Recall from Section III-E that nodes may join or leave the participant set at the beginning of each epoch. LinBFT follows the common practice (e.g., as in [18]) that the current participant set (e.g., at Epoch $e$) collectively decide on the next participant set at Epoch $e+1$. Within Epoch $e$, two tasks need to be performed regarding participant set updates. First, the nodes need to reach consensus on what are the updates to the current participant set in the next epoch. Otherwise, at Epoch $e+1$, there can be a fork on the participant set, i.e., different nodes may have conflicting views of the updated participant set. In LinBFT, this may lead to perpetual leadership contention, meaning that different honest nodes disagree on the identity of the leader, defined in Equation (2), which violates liveness.

To address this problem, in LinBFT, each update to the participant set (i.e., a node join or leave request) is simply treated as a transaction, which will be included in a new block to be added to the blockchain. The rationale is that since LinBFT guarantees deterministic safety and liveness, there must be deterministic consensus over the next participant set. On the other hand, reaching consensus in the presence of malicious nodes is essentially a BFT problem, for which LinBFT itself is the best known solution with deterministic security guarantees. Specifically, at the end of Epoch $e$, the set of join/leave requests contained in finalized blocks determine the changes in the participant set in Epoch $e+1$. In the worst case, these involve $O(n)$ join/leave transactions, e.g., when the entire participant set is replaced. Since each block contains a constant number of transactions, an epoch needs to contain $\Omega(n)$ blocks, which agrees with the assumption in Section III-E.

Second, LinBFT needs to run a DKG protocol to generate new public / private key pairs for each participant in Epoch $e+1$, which are required for creating threshold signatures. Note that LinBFT does not update keys incrementally, i.e., the key pair of a staying participant from Epoch $e$ is still generated from scratch. This is because updating keys for threshold signature is tricky in general: for instance, consider the situation with (i) an existing participant set with $n = 10$ and $f = 3$, and (ii) 3 new participant entering the system, increasing $n$ to $n' = 13$ and $f$ to $f' = 4$, respectively. Here, the old and new thresholds are $t = 2f = 6$ and $t' = 2f' = 8$, respectively. Using a DKG algorithm based on Pederson's verifiable secret sharing [17], the new and old secret keys are correlated with the new one adding two (i.e., $t'-t = 2$) more secret terms. However, these new secrets can be derived by the adversary controlling $f \geq 2$ malicious nodes.

LinBFT generates key pairs using the DKG algorithm in [7], which requires $O(n \log^3 n)$ transmissions, and provides a probabilistic guarantee on the correctness of the keys (i.e., threshold signatures can be successfully created with these keys), where the probability of failure can be made arbitrarily small, e.g., below $\rho = 10^{-18}$ as discussed in the previous subsection. This cost amortizes to $O(polylog\ n)$ over $\Omega(n)$ blocks in an epoch, which is negligible in an asymptotical sense.

In the unlikely case that the DKG algorithm fails, LinBFT still guarantees safety and liveness, as follows. Once a leader (say, $L_i$) finds that it cannot derive a threshold signature from $2f+1$ signatures, it simply broadcasts the original $2f+1$ signatures without aggregating them. The protocol then proceeds without using threshold signatures, as in PBFT, which satisfies safety and liveness. At the next block height, nodes re-run DKG to generate new key pairs for threshold signatures.

### D. Protocol Summary, Analysis, and Discussion

The proposed protocol, as presented in the previous three subsections, is basically PBFT/LVC [1] with the following modifications:

- Replace round-robin leader rotation with random leader selection according to Equation (2).
- Replace the all-to-all broadcast of *Prepare* messages with an all-to-leader gather operation. The leader then derives the corresponding commit certificate $CC_i$, defined in Equation (1), and broadcasts $CC_i$.
- Replace the all-to-all broadcast of *Commit* messages with all-to-leader gather. The leader then derives threshold signature $ts(CC_i)$, and broadcasts it.
- Apply a speculative, $O(\log n)$-time run in the ordinary case (Section IV-A). Fall back to the main protocol when the speculative run fails.
- Fall back to all-to-all broadcasts in case of a DKG failure (Section IV-C).

The following lemmas summarize LinBFT's security and performance guarantees. The proofs follow directly from those of PBFT's, and are omitted for brevity. They will appear in an extended version of the paper, if requested by the readers.

**Lemma 1 (safety).** Given $n \geq 3f+1$, after LinBFT terminates, a unique block is finalized and added to the blockchain.

**Lemma 2 (liveness).** In a partially synchronous network, given $n \geq 3f+1$, LinBFT terminates within finite time.

**Lemma 3 (linear complexity).** In a partially synchronous network, unless with negligible probability, LinBFT terminates after amortized-$O(n)$ transmissions given that (i) given $n \geq 3f+1$ and (ii) at most $O(1)$ of nodes are slashed, where the amortization is done over an epoch of $\Omega(n)$ blocks.

Finally, a public blockchain needs to reward participants in the BFT protocol as an incentive for verifying transactions. For example, the reward can be tokens in the form of transaction fees. One potential issue in LinBFT is that the block verification information, i.e., threshold signatures, do not contain the identities of the signers. Consequently, it is possible that a "lazy" node does not perform actions required by the protocol, and yet still gets rewarded for being in the participant set, which is unfair. A separate governance mechanism is required to address this issue: for instance, a leader could include in its proposed block the identities of nodes that are least active, and, thus, should be evicted from the participant set in the next epoch. Governance is an orthogonal issue that falls outside the scope of this paper, and is left as future work.

## V. RELATED WORK

BFT protocols have been extensively studied in the traditional distributed systems setting, where a client submits transactions to be verified to a cluster of $n$ nodes, among which $f$ (satisfying $n \geq 3f+1$) can be malicious. This is similar to verifying a single block in LinBFT's setting, with the notable difference that the latter does not involve a client as a distinct entity. In the literature, an early influential work is the DLS protocol by Dwork et al. [9], which achieves safety and liveness, at the expense of $O(n^4)$ communication cost, which is prohibitively high for a large cluster. A decade later, Castro and Liskov propose Practical BFT (PBFT) [5], which satisfies safety and liveness in a synchronous network with $O(n^2)$ transmissions, as long as no leader failure occurs. As pointed in [1], this is essentially an optimistic run, and the protocol falls back to DLS when the optimistic run fails. Later work, e.g., Zyzzyva [19], further improves the efficiency of the optimistic run.

In traditional BFT research discussed above, it is commonly assumed that there is a fixed cluster of verifier nodes. Meanwhile, in many protocols, the same node stays as the leader unless a view change occurs. Further, an honest node is assumed to stay honest, regardless of the number of transaction batches it verifies. With these assumptions, the traditional setting is radically different from that of that of LinBFT. For blockchains, a popular deployed protocol is Tendermint [2], which is based on PBFT and runs consensus for each block with a rotating leader scheme, as in LinBFT. Tendermint incurs $O(n^2)$ communications in the ordinary case, and $O(n^3)$ with cascading leader failures.

Recently, Casper [3] amortizes the cost of its BFT protocol by running consensus once for multiple (100 in [3]) block heights. This design, however, gives much power to the block proposer, which might lead to selfish mining. Hence, Casper involves a PoW mechanism for leader selection, which runs the risk of forks and 51% attacks. Hot-Stuff [1] improves the worst case communication complexity to $O(n^2)$, using a combination of linear view change and threshold signatures. SBFT [15] reduces the communication complexity of the ordinary case using threshold signatures and collectors. Omniledger [21] achieves $O(\log n)$ time in the best case, with the help of the CoSi protocol [25]. Nevertheless, to the author's best knowledge, LinBFT is the first protocol that achieves worst-case linear communication complexity.

Another promising trend is BFT protocols with probabilistic guarantees on safety and liveness. For instance, Dfinity [18] uses a sample set of nodes to verify a block, where the sample is obtained through a random beacon powered by VRFs. There are two issues with this approach. First, since the safety guarantee is probabilistic, the sample set needs to be sufficiently large to obtain a low probability of failure (note that a blockchain system runs indefinitely, and involves an unbounded number of blocks). This means that Dfinity still needs a scalable, deterministic sub-protocol for the sample set. Second, since the sample set is usually $o(n)$, which can be smaller than $f$, an adaptive adversary can corrupt them all. Algorand [16] addresses the latter problem with a probabilistic protocol that is robust against a powerful adversary, who is not only adaptive but has a strong influence over the network. However, as pointed out by Chan et al. [6], Algorand assumes the existence of a public key infrastructure, and is based on the assumption that a node can erase its own memory before it is corrupted by the adversary; further, the mighty adversary in Algorand needs to have a critical handicap for the protocol to work: that it cannot stop a message sent from an honest node to reach its recipients. LinBFT does not consider an adaptive adversary since (i) for a fast protocol, compromising nodes adaptively within one protocol run is rather difficult, and (ii) the only part of the protocol that is vulnerable to an adaptive adversary is random leader selection (Section IV-B), for which the adversary can break the probabilistic guarantee on $O(1)$ leader rotations by corrupting a considerable portion of the nodes; this is difficult, however, for a larger $n$.

Finally, several protocols distinguish malicious nodes who actively attack the protocol with falsified messages from ones that may fail-top (e.g., in [15]) or those that may go offline [24], and obtain stronger robustness in a setting where only a small fraction of nodes are actively malicious. LinBFT follows the more common honest/malicious participant model, and relies on an external governance module to disincentivize passive failures, as discussed in Section IV-D.

## VI. CONCLUSION

The paper proposes LinBFT, the first BFT protocol that achieves amortized $O(n)$ communication cost, while satisfying deterministic guarantees on safety and liveness, as well as several properties desirable in an open, public blockchain setting. An immediate future direction is to implement LinBFT as a deployable prototype, and benchmark it against existing solutions. Additionally, it is interesting to investigate the combination of LinBFT with a randomized BFT protocol, as well as other scaling options such as sharding [21].